\documentclass[prl,twocolumn,superscriptaddress,aps,floatfix,nofootinbib]{revtex4-2}

\usepackage{braket}
\usepackage{graphicx}
\usepackage[normalem]{ulem}
\usepackage{amsmath}
\usepackage{mathrsfs}
\usepackage[dvipsnames]{xcolor}
\usepackage{longtable}
\usepackage{amsfonts}
\usepackage{tikz}
\usepackage{float}
\usepackage{physics}
\usepackage[colorlinks,linkcolor=blue,citecolor=blue,anchorcolor=blue]{hyperref}
\usepackage{bm}
\usepackage{enumitem}
\usepackage{verbatim}
\usepackage{orcidlink}
\usepackage{xcolor}

\begin{document}	
\title{Non-destructive cavity readout of molecules for precision measurements}

\author{Alejandro Salas-Estrada$^{1,^\ast}$
        Silviu-Marian Udrescu$^{2,3,^\ast}$\orcidlink{0000-0002-1989-576X}, Geoffrey Zheng$^{3}$\orcidlink{0000-0002-9312-0102}, Qian Wang$^{3}$\orcidlink{0000-0003-2853-8534}, 
        Arian Jadbabaie$^{4,5}$, Vladan Vuleti\'{c}$^{4,5}$\orcidlink{0000-0002-9786-0538}, 
       David DeMille$^{2,3,6}$\orcidlink{0000-0001-7139-4121}, Ronald F. Garcia Ruiz$^{4,5}$\orcidlink{0000-0002-2926-5569}, Edwin Pedrozo-Pe\~{n}afiel$^{1,^\dagger}$\orcidlink{0000-0003-4348-9910} \\~\\
\small \textit{$^{1}$Department of Physics, University of Florida, Gainesville, FL 32608, USA}\\
\small \textit{$^{2}$Department of Physics and Astronomy, Johns Hopkins University, Baltimore, MD 21210, USA}\\
\small \textit{$^{3}$Department of Physics, University of Chicago, Chicago, IL 60637, USA}\\
\small \textit{$^{4}$Department of Physics, Massachusetts Institute of Technology, Cambridge, MA 02139, USA}\\
\small \textit{$^{5}$MIT-Harvard Center for Ultracold Atoms and Research Laboratory of Electronics}\\
\small \textit{$^{6}$Physics Division, Argonne National Laboratory, Lemont, IL 60439, USA}\\
\small $^\dagger$Corresponding author: 
\href{mailto:epedrozo@ufl.edu}  {epedrozo@ufl.edu} \\~\\
\small $^\ast$These authors contributed equally to this work.}

\begin{abstract}

\bf We propose a non-destructive method to measure the population of molecules in a selected rotational-hyperfine state by coupling them to a high-finesse optical cavity. In contrast to traditional techniques, our approach enables fast (less than 1 ms) repeated measurements with reduced heating and losses, and with precision below the standard quantum limit. The method is particularly advantageous for radioactive molecules, systems of high interest for symmetry violation searches, for which production and sample size are limited, and repeated interrogation is essential for improved sensitivity.

\end{abstract}

\maketitle

\emph{Introduction}---During the past decades we have witnessed remarkable progress in understanding the laws of nature, driven by a synergy between increasingly precise experiments and theoretical developments \cite{navas2024review}. However, many observed phenomena are still evading a comprehensive explanation, such as the nature of Dark Matter and Dark Energy or the origin of the matter-antimatter asymmetry of our visible Universe. 

Recently, precision measurements using molecules have emerged as a fruitful approach to trying to answer many of these questions, through searches for violations of the fundamental symmetries of nature, such as parity (P) and time-reversal (T) \cite{safronova2018search}. Diatomic molecules have already set the most stringent bounds on the P,T-odd electron electric dipole moment (EDM) \cite{roussy2023improved,acme2018improved}, while various diatomic and polyatomic molecules are currently being used for searches of the P,T-odd nuclear Schiff moment and magnetic quadrupole moment \cite{hutzler2020polyatomic,grasdijk2021centrex,arrowsmith2024} or the P-odd, T-even nuclear anapole moment \cite{altuntacs2018demonstration,blanchard2023using,karthein2024electroweak}. Such experiments are expected to provide more than three orders of magnitude enhanced sensitivity to the sought-after signals compared to atoms \cite{safronova2018search}.

A further enhancement in sensitivity to nuclear symmetry-violating effects, of up to three orders of magnitude, is expected from using molecules containing short-lived, octupole-deformed nuclei \cite{yang2023laser,arrowsmith2024,engel2025nuclear}. The recently achieved spectroscopic investigation of such species opened the way for using radioactive molecules for future precision measurements \cite{garcia2020spectroscopy,udrescu2021isotope,udrescu2024precision,wilkins2025observation,wilkins2026ionization,conn2025production}. However, despite this progress, working with radioactive molecules poses significant challenges \cite{arrowsmith2024}. In particular, given their short lifetimes and low natural abundances, they can often be produced only in minuscule amounts \cite{garcia2020spectroscopy,udrescu2021isotope,udrescu2024precision,wilkins2025observation}. This requires the development of novel techniques that can allow high-duty cycles in the usage of the available molecules, as well as fast experimental protocols \cite{arrowsmith2024,udrescu2024precisionPRR,karthein2024electroweak}.

A promising route to harnessing the power of radioactive molecules, facilitating long coherence times and good control over various sources of systematic uncertainties is the use of ultracold molecular techniques \cite{arrowsmith2024,demille2024quantum, langen2024quantum}. Major advances in cooling, trapping, and coherent control of stable molecules have recently been achieved~\cite{barry2014magneto,burau2023blue,truppe2017molecules_tarbutt,padilla2025_AlF_mot,park2023magnetic,miller2024two,bigagli2024observationBECmolecules_will,picard2025entanglement,christakis2023probing,CheukCaFRotationalEntanglement,vilas2024optical,ruttley2025long_cornish}. However, a major challenge in existing experimental protocols is the readout of the internal molecular state from which the signal of interest is extracted. Current approaches typically rely on destructive or strongly scattering-based detection, either by dissociating the molecule into constituent atoms~\cite{christakis2023probing,park2023magnetic,miller2024two,picard2025entanglement,kondov2019molecularclock_zelevinsky} or by fluorescence imaging~\cite{CheukCaFRotationalEntanglement,vilas2024optical,highfidelityflour2021_mccarron}. In particular, low-backaction readout techniques enabling repeated interrogation of the same molecular ensemble with reduced heating and state redistribution are highly desirable.

These processes can be slow, while leading to substantial molecular losses, thus representing a significant challenge, especially when working with radioactive and low-abundance molecular species~\cite{arrowsmith2024}. In this letter we propose a novel approach for non-destructively measuring the internal molecular states using the cavity quantum electrodynamics (cavity-QED) platform~\cite{kimble1998strong, haroche2013nobel}, which can overcome these limitations and is particularly well suited to the stringent requirements imposed by radioactive and low-abundance molecular species~\cite{arrowsmith2024}.

While our method is general, we will focus on a particular situation of interest for future experimental searches of P-odd, T-even and P,T-odd effects. We assume that the molecules are initially prepared in an equal superposition of two internal states and then allowed to evolve for a fixed period of time under the influence of the symmetry-violating interaction of interest, typically in the presence of applied electric and magnetic fields. The accumulated phase information is encoded in the final state populations. By measuring the final population in the two levels, we can extract the desired signal with a statistical uncertainty limited by the quantum projection noise, also known as the standard quantum limit (SQL), which scales with the number of interrogated molecules, $N$, as $\frac{1}{\sqrt{N}}$. However, in order to reach this limit, we need to be able to count the number of molecules in a given internal state with a variance below the SQL. Otherwise, the measurement would be limited by the readout capabilities of our setup and not by the fundamental limit due to the probabilistic nature of quantum mechanics. 

Therefore, our goal is to perform the readout step of our experiment with a variance below the SQL. We also aim to achieve this as quickly as possible and with a minimal loss of molecules, which would allow significantly higher duty cycles compared to available methods. This is particularly relevant for measurements performed online at radioactive beam facilities, where the allocated time for individual experiments is limited, usually on the order of a few weeks per year \cite{yano2007radioactive,glasmacher2017facility,borge2017isolde}. 

\begin{figure}[]
    \centering
    \includegraphics[width=\columnwidth]{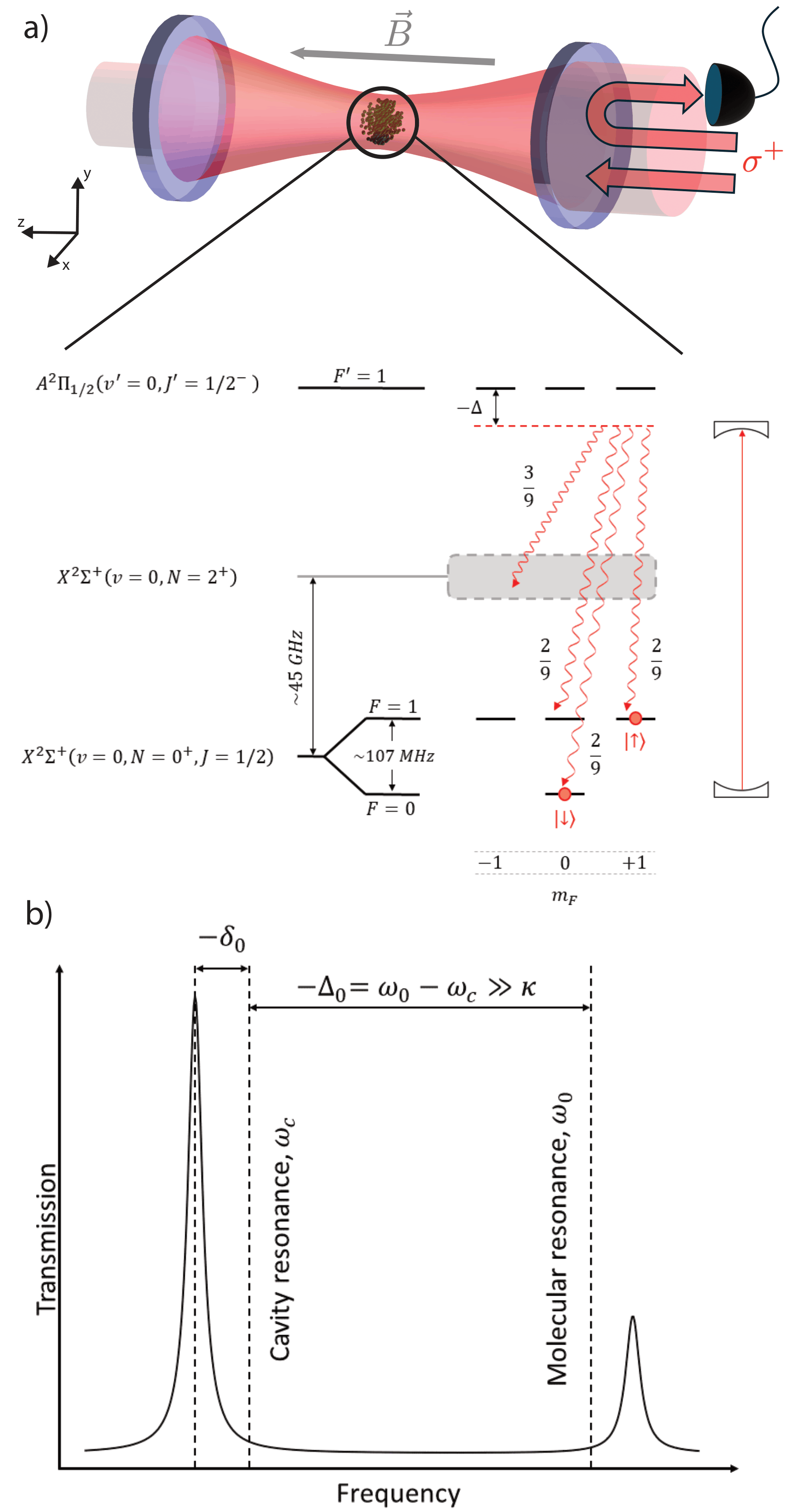}
    \caption{Molecules interacting strongly with the light inside a high finesse optical cavity (a) Top: The molecules are optically trapped (trapping light not shown) close to the beam waist of the light building up inside the cavity (red). An externally applied magnetic field defines the quantization axis along the cavity (gray arrow). A photodetector captures the reflected light, which provides a dispersive measurement of the molecular state population. Bottom: Relevant energy levels in SrF. The transition $\ket{\downarrow} \rightarrow \ket{e}$ is driven using off resonant light (red arrow) with a frequency close to the dressed cavity resonance. The light detuning from the molecular transition is $\Delta \gg \Gamma$. The possible molecular decay pathways from $\ket{e}$ and their probabilities are shown (wavy arrows). (b) For a cavity far-detuned from the molecular resonance, $\omega_0-\omega_c\gg \kappa$, the cavity transmission spectrum displays two asymmetrical peaks. The more prominent one (left), which is close to the resonant frequency of the cavity, shifts proportionally to the number of molecules coupled to the cavity, by an amount given by $\delta_0 = N_{\downarrow}\eta\Gamma\kappa/4\Delta_0$. See main text for details.}
    \label{fig:cavity_shift}
\end{figure}

Our proposed method relies upon engineering strong couplings between the internal molecular degrees of freedom and the light inside a high-finesse optical cavity. This research direction proved extremely fruitful in atomic systems, beginning with pioneering single-atom cavity QED experiments~\cite{Kimble1992,pinkse2000trapping,Fortier_singleatomsCQED_Chapman_2007,terraciano2009photon}, followed by collective measurements and entanglement generation in atomic ensembles~\cite{cavityfeedbackSq_Schleier-Smith2010,Cox_RecordSqueezing,hosten2016measurement,pedrozo2020entanglement,periwal2021programmable,robinson2024directcomparisonsq_yegroup,Shadmany_highNA_simongroup_2025}, and later by the exploration of cavity-mediated many-body physics in quantum-degenerate gases~\cite{brennecke2007cavity_BEC_esslinger,Kroeze_spinorBEC_20218_Lev,roux2020stronglyEPFL}. Extending these techniques to molecules did not seem possible until recently. This was mainly due to their much more complicated energy-level structure compared with atoms, as well as the difficulty of reaching temperatures on the order of tens of $\mu$K that are suitable for cavity QED techniques~\cite{lev2008prospects}. However, the advances in ultracold molecule research in recent years are now making it possible to overcome these longstanding barriers \cite{lu2024ramancooling_cheuk,DeMilleSrFCollisions}. Thus, adding cavity-QED methods, in particular non-destructive cavity readout, to the repertoire of tools available in the study of various molecular species of interest appears within reach. For concreteness, in the remainder of this paper we present a feasible implementation of internal-state cavity readout using SrF molecules. However, the results are general and applicable to a wide range of stable and radioactive molecules of interest \cite{arrowsmith2024,demille2024quantum, langen2024quantum}.

\emph{Experimental Proposal}---The SrF molecules are produced inside a He-filled buffer gas cell at $\sim 4$~K. The molecules leaving the cell, with forward velocities of $\sim 140$ m/s, are slowed down using the white light slowing technique and then captured in a Magneto-Optical Trap (MOT) \cite{barry2014magneto,DeMilleSrFCollisions}. Here, they are cooled down in several stages to $\sim 10$ $\mu$K \cite{DeMilleSrFCollisions}. At this point we expect to have $10^4-10^5$ molecules trapped in a MOT with $\sim 100$ $\mu$m diameter (see Ref. \cite{DeMilleSrFCollisions} and Supplemental Material \cite{supp} for details\nocite{graceli2025CBMOTTheory, hallas2026CaOH_CBMOT_published, scarlett2024CaF_CBMOT, zeng2026bafbluemot_published, SrOHODTandCBMOTPublished, QianWangSrFThesis, scarlett2024CaF_CBMOT, TarbuttCaFMagTrapMWTransfer, DoyleCaFRotationalEntanglement, DoyleCaOHQuantumStateControl, EricNorrgardPhDThesis, SrFSpectroscopy1970s, VarunJorapurThesis}). The molecular cloud will then be superposed with the optical lattice created by a standing wave inside the optical cavity. This lattice will be used as a conservative trap for the molecules while ensuring that they are located close to the cavity waist. We expect a transfer efficiency from the MOT to the cavity lattice of about $1-10\%$. Then, the measurement protocol for the quantity of interest (e.g. symmetry violating moments) will begin. For most ongoing and planned experiments, at the end of this step the molecules will be in an almost equal superposition of two internal states. The signal we want to measure is proportional to the population difference between these two states, which we aim to extract using our cavity-based protocol.

However, in general, the internal states used for spectroscopy or phase accumulation are not optimal for performing population counting using cavity readout. Therefore, the molecular population may first need to be transferred to states optimized for detection. The levels used for readout are selected to minimize the variance in the inferred molecular number, and they are located $\sim 100$'s MHz - $10$'s GHz away from the states used during the actual measurement, depending on the chosen experimental protocol \cite{nl2018measuring,verma2020electron,wilkins2025observation,arrowsmith2024}. Such population transfer between the spectroscopy states and the states optimized for cavity readout can be implemented using the appropriate microwave or RF coupling scheme for the chosen molecular levels, with high efficiency (see Supplemental Material \cite{supp} for details), while preserving the coherence and population difference between the two levels under consideration. In our implementation, using SrF molecules, the states used for readout are chosen to be $\ket{\downarrow} = \ket{X^2\Sigma(\nu=0,N=0^+,J=1/2,F=0,m_F=0)}$ and $\ket{\uparrow} = \ket{X^2\Sigma(\nu=0,N=0^+,J=1/2,F=1,m_F=1)}$, as shown in Fig.~\ref{fig:cavity_shift}(a). $X^2\Sigma$ denotes the ground electronic state of the molecule, $\nu$ is the molecular vibrational quantum number, $\textbf{N
}$ is the total angular momentum excluding the electronic and nuclear spin; $\textbf{J}$ and $\textbf{F}$ are the total angular momentum including electronic and nuclear spin, respectively, and $m_F$ is the magnetic quantum number associated with $\mathbf{F}$.

For the readout step, the quantization axis, $\mathbf{\hat{z}}$, is defined using a magnetic field, $B\approx 1$ G, oriented along the cavity. A probe light beam with $\sigma^+$ polarization and frequency $\omega_l$, propagating along the cavity axis as shown in Fig.~\ref{fig:cavity_shift}(a), will be nearly resonant with a cavity mode at frequency $\omega_c$, and detuned by $\Delta = \omega_l-\omega_0 \gg \Gamma$ from the molecular transition $\ket{\downarrow}\leftrightarrow \ket{e}$, with $\ket{e} = \ket{A^2\Pi_{1/2}(\nu'=0,J'=1/2^{-},F'=1,m_{F}'=1)}$. Here $\hbar\omega_0$ is the energy difference between $\ket{\downarrow}$ and $\ket{e}$ and $\Gamma/2\pi \approx 7~\mathrm{MHz}$ is the total decay rate of the excited state, $\ket{e}$. The probe light power will be chosen such that we operate in the low saturation regime, minimizing heating from spontaneous emission and thus the probability for the molecules to escape from the optical trap, thus addressing a key limitation of traditional molecule counting methods~\cite{vilas2024optical,highfidelityflour2021_mccarron,shuman2009force}.

In the absence of molecules, the cavity reflection spectrum exhibits an anti-resonance centered at $\omega_c$ with linewidth $\kappa$. When molecules in the state $\ket{\downarrow}$ are present, they modify the optical properties of the intracavity medium. In the dispersive regime we are working in, this leads to a shift of the cavity resonance that can be understood as a collective, state-dependent change in the effective index of refraction experienced by the light \cite{tanji2011classical}. The shift of the cavity mode scales with the number of molecules in $\ket{\downarrow}$, $N_{\downarrow}$, and is given by 
\begin{equation}
    \delta_0 = \frac{N_{\downarrow}\eta\Gamma\kappa}{4\Delta_0},
    \label{cavity_shift}
\end{equation}
where $\Delta_0=\omega_c-\omega_0$ is the cavity-molecule detuning and $\eta$ is the single-molecule cooperativity, defined as \cite{duan2020thesis}:
\begin{equation}
    \eta = \frac{24\mathcal{F}}{\pi k^{2}w_{0}^2}\frac{\Gamma_{e\downarrow}}{\Gamma},
    \label{eta}
\end{equation}
with $\mathcal{F}$ being the finesse of the cavity, $w_{0}$ the size of the cavity waist, $\Gamma_{e\downarrow}$ the partial decay rate of $\ket{e}$ into $\ket{\downarrow}$, and $k=\omega_{l}/c$. By scanning the laser across the shifted cavity peak, we can determine $N_{\downarrow}$ with minimal losses and heating of the molecular ensemble \cite{zhang2012unitresolution}. Note that, given the $\sigma_{+}$ polarization of the probe light and angular momentum selection rules, the $\ket{\uparrow}$ state has a much smaller coupling to the intracavity field and thus it won't significantly influence the shift of the cavity resonance peak. However, the population in the $\ket{\uparrow}$ state, $N_\uparrow$, can still be extracted by applying a resonant $\pi$-pulse between $\ket{\downarrow}$ and $\ket{\uparrow}$ and repeating the counting procedure. For the rest of the paper, we quantify the population difference between the two levels considered as $J_z \equiv \left(N_\uparrow - N_\downarrow\right)/2$.

\emph{Numerical Simulations}---The minimum variance in estimating $J_z$ is determined by the cavity parameters, molecular properties, and the molecule-light interaction strength.
It can be shown that \cite{chen2014cavity,li2022collective, zhang2012unitresolution}:
\begin{equation}
    \frac{(\Delta J_z)^2}{(\Delta J_z)^2_{\mathrm{SQL}}} = \frac{1}{N}\bigg(\frac{4}{n_{s}q\widetilde{f}_{meas}} + 4p_{\mathrm{flip}}n_{s} + p_{\mathrm{loss}}n_{s}\bigg)
    \label{eq:J_sq}
\end{equation}
where $(\Delta J_z)^{2}$ is the variance in the measurement of $J_{z}$, $(\Delta J_z)^{2}_{SQL}=(N_{\downarrow}+N_{\uparrow})/4 = N/4$ is the standard quantum limit variance, $n_{s}$ is the number of photons scattered into free space, $q$ is the detector's quantum efficiency, {$\widetilde{f}_{meas}$ is the Fisher information in the reflected light per scattered photon}, $p_{\mathrm{flip}}$ is the probability that a molecule in the $\ket{\downarrow}$ state will flip to $\ket{\uparrow}$ after scattering a photon, and $p_{\mathrm{loss}}$ is the probability that the molecule will go to any other internal state after a photon scattering event (see Supplemental Material \cite{supp} for details). Thus, minimizing the variance requires finding an optimal balance between the first term, which models the photon shot noise reduction as more photons are detected, and the last two terms modeling the Raman scattering noise \cite{chen2014cavity}.

The value of $p_{\mathrm{flip}}$ and $p_{\mathrm{loss}}$ are fixed and given by the vibrational and rotational branching ratios \cite{lev2008prospects}. In the specific case of SrF, for the chosen $\ket{\downarrow}$, $\ket{\uparrow}$, and $\ket{e}$ states, the branching probabilities are $p_{\mathrm{flip}}=2/9$ and $p_{\mathrm{loss}}=5/9$~\cite{barry2014magneto}, as determined by angular momentum selection rules~\cite{wall2008lifetime}. Vibrational branching is negligible due to the highly diagonal Franck–Condon factors of SrF~\cite{barry2014magneto}. As a result, free-space scattering can rapidly redistribute population across many internal states from which molecules cannot be efficiently recovered~\cite{shuman2009force}. However, a far-detuned cavity configuration, such as the one we propose, reduces the free-space scattering~\cite{chen2014cavity}, $n_s$, while allowing useful information about the ensemble to be extracted from the reflected or transmitted photons~\cite{zhang2012unitresolution}.

The information carried by the detected photons, and hence the reduction of the shot noise in estimating $J_z$, is quantified by the Fisher information $F$~\cite{li2022collective}, proportional to the single-molecule cooperativity, $\eta$ (see Supplemental Material \cite{supp} for details). As seen in Eq. \ref{eta}, increasing $\eta$ requires high finesse and a small cavity waist. Therefore, for the numerical simulations presented herein, we consider a near-concentric cavity with the mirrors radius of curvature $R=2.5$ cm and a length $L$ set $20$ $\mu$m shorter than the concentric limit. The cavity uses a low-reflectance input mirror and a second mirror with near-unity reflectance, forming an effectively single-sided cavity. For our probing wavelength of $\lambda \approx 663$ nm, this geometry yields a cavity waist $w_0 \approx 10$ $\mu$m. Even with a modest finesse $\mathcal{F}=5,000$, this results in $\eta>1$, known as the strong coupling regime, where emission into the cavity mode exceeds free space emission~\cite{tanji2011classical}. We assume a detuning between the cavity mode and the molecular transition of $\Delta_0 = \omega_c - \omega_0 \approx - 1$ GHz. This optimal value balances the need to work in the low saturation regime, which requires a larger detuning for a fixed laser power, with maintaining a strong signal in the measured spectrum, requiring a smaller $\Delta$ (Eq. \ref{cavity_shift}).
For the simulations, we assume a total detection efficiency $q=0.4$. The probe is assumed to be locked to the dressed-cavity resonance using a phase-sensitive readout of the reflected light, e.g., Pound-Drever-Hall (PDH) signal~\cite{drever1983laser}. In practice, the resonance position can be located with a brief low-power adaptive search before the measurement~\cite{MatthewPeters_AdaptivecQEDMeasurement}.

\begin{figure}[]
    \centering
    \includegraphics[width=0.95\columnwidth]{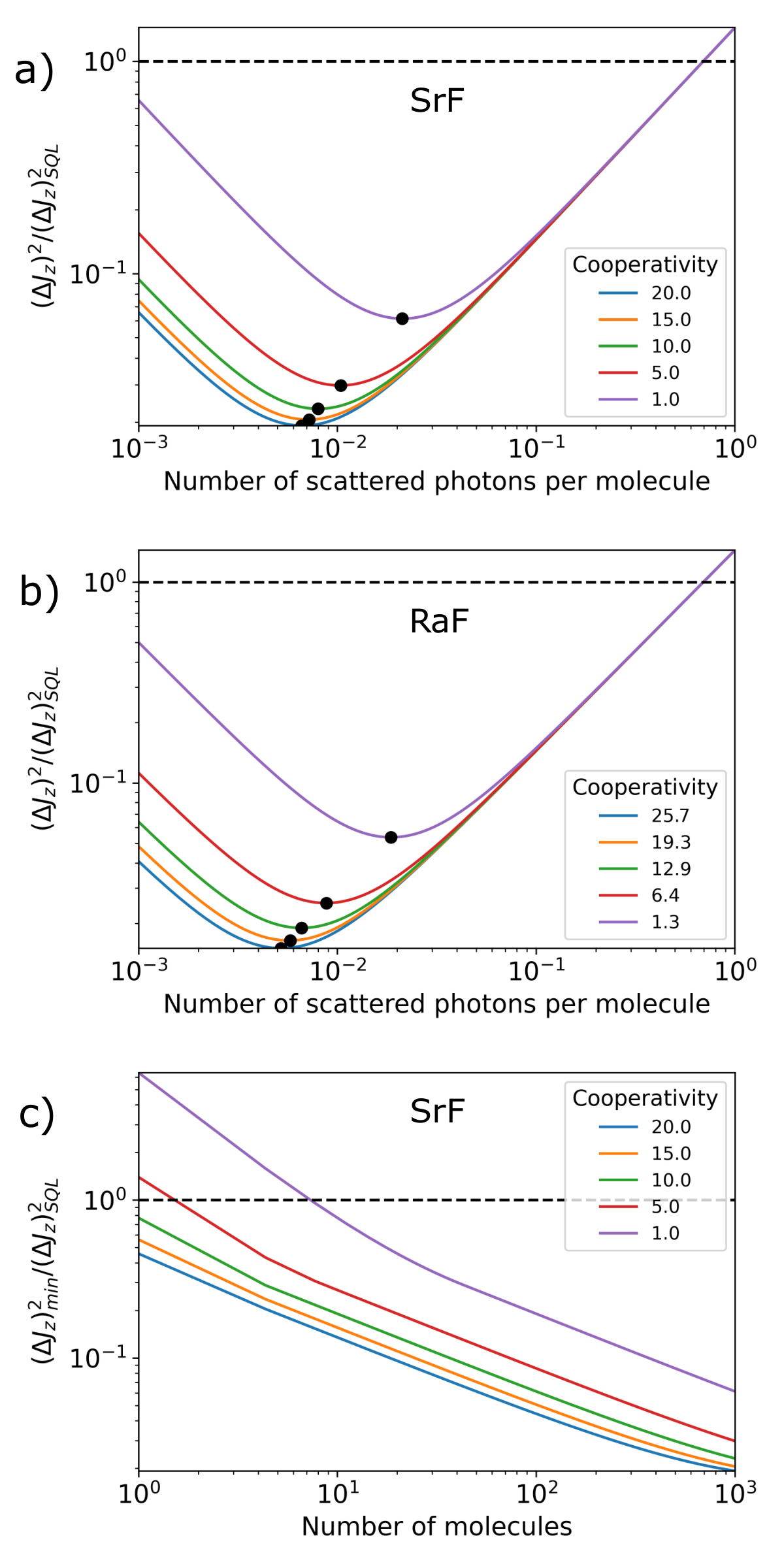}
    \caption{\textbf{(a)} Measurement variance normalized to the SQL as a function of the number of scattered photons per molecule for an ensemble of $N = 1000$ SrF molecules with detection efficiency $q = 0.4$. Each curve corresponds to a cavity of length $\sim 5\,\mathrm{cm}$, mode waist $10\,\mu\mathrm{m}$, and detuning $1\,\mathrm{GHz}$ from the molecular resonance. The finesse ranges from $\sim 7{,}000$ to $140{,}000$, determining the cooperativity and linewidth. Black points indicate the optimal measurement positions where Raman noise equals photon shot noise. The dashed line marks the SQL. The corresponding measurement time is on the sub-millisecond scale. 
\textbf{(b)} Same as (a) for RaF, assuming identical ensemble size and cavity parameters. 
\textbf{(c)} Minimum measurement variance as a function of the number of SrF molecules in the cavity. The dashed line marks the SQL. Similar results for RaF are shown in the Supplemental Material \cite{supp}.}
    \label{fig:DJ}
\end{figure}

\emph{Results}---As shown in Fig. \ref{fig:DJ} (a), for the SrF the measurement variance $(\Delta J_z)^2$ follows the behavior predicted by Eq. (\ref{eq:J_sq}) for different values of the single-molecule cooperativity $\eta$. Initially, the variance decreases due to the increase in total Fisher information carried by the photons transmitted through the cavity. At larger probe powers, the variance increases as Raman noise from free-space scattered photons becomes dominant~\cite{chen2014cavity, li2022collective}. For a broad range of cooperativities, $1 \lesssim \eta \lesssim 20$, the minimal achievable variance lies below the SQL, $(\Delta J_z)^2_{\mathrm{SQL}}$.

To explore the applicability of this approach to molecules relevant to fundamental symmetry tests, we also analyzed the case of $^{226}$RaF. 
Radium-containing molecules are expected to play a central role in future searches for violations of fundamental symmetries~\cite{arrowsmith2024,jadbabaie2026_radioactivemolecules}. We assume cavity parameters identical to those used for SrF and choose the same $\ket{\downarrow}, \ket{\uparrow}, \ket{e}$ states, since the spin-rotation and hyperfine structures of $^{88}$SrF and $^{226}$RaF are identical, apart for the energy splitting between levels. 

Because the wavelengths of their cycling transitions differ, Eq. (\ref{eta}) predicts different cooperativities for SrF and RaF.
As shown in Fig. \ref{fig:DJ} (c), the proposed cavity geometry enables precision below the SQL for a wide range of cooperativities and molecule numbers. As in the SrF case, the measurements can be performed on a fast time scale, on the order of sub-milliseconds  while reducing the molecular number by $\sim 1\%$, or less for $\eta\gtrsim5$.
Finally, Fig. \ref{fig:DJ} (c), shows the minimal $(\Delta J_z)^2$ value, as a function of the total number of SrF molecules, $N = N_\downarrow+N_\uparrow$. 
For the cavity parameters considered here, the measurement variance remains below the SQL across the entire range of ensemble sizes studied. These results indicate that the proposed readout scheme is robust over a wide range of experimentally relevant molecule numbers. Similar behavior is obtained for RaF (see Supplemental Material \cite{supp}).
A key advantage of this approach, compared with conventional readout techniques that typically remove molecules from the trap, is that the measurement is nearly nondestructive. As a result, the experimental sequence can be restarted using the same trapped molecular ensemble once $J_z$ is measured. 

For the parameters considered here, Raman-induced loss during the measurement is small and does not set the dominant limit on ensemble reuse (see Supplemental Material~\cite{supp}). In the absence of additional technical limitations, the low free-space scattering rate would in principle allow hundreds of repeated measurements on the same trapped ensemble before the molecular number is reduced to $1/e$ of its initial value (see Supplemental Material~\cite{supp}). In practice, however, the number of useful interrogation cycles is expected to be primarily limited by the efficiency of state reinitialization. For realistic SrF state-preparation efficiencies ($\sim90\%$ per cycle), we estimate that the same ensemble could support on the order of ten repeated interrogation cycles before reloading becomes necessary. This limitation could be substantially mitigated using cavity-assisted optical pumping schemes~\cite{morigi2007cavity}.

\emph{Conclusion}---We have presented a cavity-based readout scheme for internal molecular quantum states that meets the stringent requirements of next-generation precision measurement experiments, particularly those involving species highly sensitive to symmetry violations effects, such as radioactive molecules \cite{arrowsmith2024}. By engineering a strong coupling between selected molecular energy levels and a high-finesse optical cavity field operated in the far-detuned, low-saturation regime, our method enables sub-SQL counting resolution on sub-millisecond timescales, with minimal molecular loss. 

A key advantage of this approach is that the measurement is nearly nondestructive, enabling higher duty cycles~\cite{Bowden_Schippo_clock_QND_2020}. This is particularly important for radioactive species at beam facilities, where both sample availability and measurement time are limited~\cite{udrescu2024precision,wilkins2025observation,arrowsmith2024}. Beyond state readout, cavity-based detection naturally enables cavity-mediated spin squeezing and other collective quantum protocols, opening a route toward entanglement-enhanced molecular sensors, with high precision in probing symmetry violations and testing physics beyond the standard model \cite{arrowsmith2024,demille2024quantum}. 
Beside precision measurements, this will open new research avenues in quantum chemistry, quantum computing, and quantum simulation with ultracold molecules \cite{langen2024quantum,karman2024ultracold,cornish2024quantum}. Combining the short-range interactions provided by the strong dipole-dipole interaction present in molecules with the long-range cavity-mediated interactions, can represent an interesting platform to engineer complex Hamiltonians for many-body physics simulations~\cite{langen2024quantum,karman2024ultracold,cornish2024quantum}.

\bigskip
\emph{Acknowledgments}---A.S.-E. acknowledges support from the University of Florida CCMS Fellowship. E.P.-P. acknowledges support from the University of Florida startup funds and the Office of Naval Research under Award No. N00014-26-1-2211. E.P.-P. thanks Simone Colombo, Zeyang Li, and Sepehr Ebadi for useful discussions. R.F.G.R and A.J. acknowledge support from the U.S. Department of Energy, Office of Science, Office of Nuclear Physics under grants DE-SC0026217.

\bibliographystyle{apsrev4-2}
\bibliography{references}

\clearpage
\onecolumngrid
\setcounter{figure}{0}
\renewcommand{\thefigure}{S\arabic{figure}}
\setcounter{table}{0}
\renewcommand{\thetable}{S\arabic{table}}
\setcounter{equation}{0}
\renewcommand{\theequation}{S\arabic{equation}}

\begin{center}
{\large\bf Supplemental Materials: Non-destructive cavity readout of molecules for precision measurements}\\[1.0em]

Alejandro Salas-Estrada$^{1,^\ast}$
        Silviu-Marian Udrescu$^{2,3,^\ast}$\orcidlink{0000-0002-1989-576X}, Geoffrey Zheng$^{3}$\orcidlink{0000-0002-9312-0102}, Qian Wang$^{3}$\orcidlink{0000-0003-2853-8534}, Arian Jadbabaie$^{4,5}$, Vladan Vuleti\'{c}$^{4,5}$\orcidlink{0000-0002-9786-0538}, 
       David DeMille$^{2,3,6}$\orcidlink{0000-0001-7139-4121}, Ronald F. Garcia Ruiz$^{4,5}$\orcidlink{0000-0002-2926-5569}, Edwin Pedrozo-Pe\~{n}afiel$^{1,^\dagger}$\orcidlink{0000-0003-4348-9910} \\~\\
\small \textit{$^{1}$Department of Physics, University of Florida, Gainesville, FL 32608, USA}\\
\small \textit{$^{2}$Department of Physics and Astronomy, Johns Hopkins University, Baltimore, MD 21210, USA}\\
\small \textit{$^{3}$Department of Physics, University of Chicago, Chicago, IL 60637, USA}\\
\small \textit{$^{4}$Department of Physics, Massachusetts Institute of Technology, Cambridge, MA 02139, USA}\\
\small \textit{$^{5}$MIT-Harvard Center for Ultracold Atoms and Research Laboratory of Electronics}\\
\small \textit{$^{6}$Physics Division, Argonne National Laboratory, Lemont, IL 60439, USA}\\
\small $^\dagger$Corresponding author: 
\href{mailto:epedrozo@ufl.edu}{epedrozo@ufl.edu} \\~\\
\end{center}

\vspace{1cm}

The theory of light-matter interactions in optical cavities has been extensively developed~\cite{li2022collective}. In particular, standard input-output treatments provide the transmitted and reflected optical fields generated when an ensemble of oscillator-like emitters is coupled to a cavity and probed through one of its mirrors. We employ this formalism to estimate the uncertainty of the population measurement described in the main text.

\section{I. Notation}\label{sec:notation}

Our notation follows Ref.~\cite{li2022collective}, with minor modifications appropriate for a molecular ensemble coupled to the cavity field. The definitions used throughout this Supplemental Material are summarized in Table~I.

\begin{table}[H]
\centering
\begin{tabular}{|l|l|}
\hline
\textbf{Symbol}                    & \textbf{Definition}                                                  \\ \hline
$\omega_0$                & Molecular transition frequency                              \\ \hline
$\omega_c$                & Bare cavity mode frequency                                  \\ \hline
$\omega_l$                & Probe laser frequency                                       \\ \hline
$\Gamma$                  & Molecular transition linewidth                              \\ \hline
$\kappa$                  & Cavity linewidth                                            \\ \hline
$x_c$                     & $x_c \equiv \frac{\omega_c - \omega_0}{\kappa/2}$               \\ \hline
$x_0$                     & $x_0 \equiv \frac{\omega_l - \omega_0}{\Gamma/2}$               \\ \hline
$\mathcal{L}_a$           & Absorptive linewidth $\mathcal{L}_a \equiv \frac{1}{1+x_0^2}$    \\ \hline
$\mathcal{L}_d$           & Dispersive linewidth $\mathcal{L}_d \equiv \frac{-x_0}{1+x_0^2}$ \\ \hline
$R_{i(o)}$                   & Reflectance of the input(output) mirror                           \\ \hline
$T_{i(o)}$                   & Transmittance of the input(output) mirror                          \\ \hline
$N$                       & Total number of molecules in the cavity                     \\ \hline
$N_{\uparrow,\downarrow}$ & Number of molecules in the $\ket{\uparrow}, \ket{\downarrow}$ state    \\ \hline
$\eta$                    & Single-molecules cooperativity                                 \\ \hline
$w_0$                     & Cavity mode waist                                           \\ \hline
$\mathcal{F}$             & Cavity finesse \\
\hline
\end{tabular}
\caption{Definitions of the mathematical symbols used throughout this Supplemental Material.}
\label{table_notation}
\end{table}

\section{II. Formalism}\label{sec:formalism}
Following the setup shown in Fig.1(a), we describe the intracavity electric field using the auxiliary quantity

\begin{equation}
    \mathcal{E} = E(0, 0, z)\sqrt{\frac{\varepsilon_0c\pi w^2(z)}{2}}
\end{equation}

where $E(0,0,z)$ and $w(z)$ denote the electric field and mode waist at position $z$, respectively. The quantity $\mathcal{E}$ is related to the optical power through $P = |\mathcal{E}|^2/2$.

The intracavity electric field is then given by

 \begin{equation}
 \label{eq:Ec}
     \mathcal{E}_{c} = \frac{\mathcal{F}}{\pi}\frac{iT_i^{1/2}}{1 + N_{\downarrow}\eta\mathcal{L}_a(x_a) - i(x_c + N\eta\mathcal{L}_d(x_a))}\mathcal{E}_{in}
 \end{equation}
 
From eq.~\eqref{eq:Ec} and the mirrors' transmittance we can find the ratio of transmitted to input power:
\begin{equation}
\label{eq:T}
    \mathcal{T} = \frac{4T_{i}T_{o}}{(T_{i}+T_{o})^2}\frac{1}{(1+N_{\downarrow}\eta\mathcal{L}_{a}(\omega_l))^{2} + \big(x_c+N_{\downarrow}\eta\mathcal{L}_{d}(\omega_l)\big)^{2}}
\end{equation}

For a cavity with identical input and output mirrors,
\begin{equation}
    \mathcal{T}_0=\frac{1}{(1+N_{\downarrow}\eta\mathcal{L}_{a}(\omega_l))^{2} + \big(x_c+N_{\downarrow}\eta\mathcal{L}_{d}(\omega_l)\big)^{2}}
\end{equation}

The ratio between the scattered and input powers is similarly given by
\begin{equation}
\label{eq:S}
    \mathcal{S} = \frac{T_{i}+T_{o}}{T_{o}}\mathcal{T}N_{\downarrow}\eta\mathcal{L}_{a}(\omega_l)
\end{equation}

By energy conservation, the ratio between the reflected and input powers is
\begin{equation}
    \label{eq:R}
    \mathcal{R} = 1 - \mathcal{T} - \mathcal{S}
\end{equation}

From these expressions, we can derive the number of photons scattered during a measurement performed at fixed probe frequency. If $P_{\mathrm{in}}$, $\omega_l$, and $\tau$ denote the probe input power, probe frequency, and measurement duration, respectively, then the number of scattered photons is

\begin{equation}
\label{eq:ns}
    n_s = \frac{P_{in}\tau}{\hbar\omega_l}\mathcal{S}(\omega_l)
\end{equation}

As discussed in the main text, molecular scattering modifies the intracavity field, such that measurements of the outgoing light provide information about the molecular state. The minimum uncertainty in a measurement of $J_z$ is then determined by the Fisher information of the detected light through the Cramér-Rao bound:

\begin{equation}
\label{eq:DJz_bound}
    (\Delta J_z)^2_{CR} = \frac{1}{\widetilde{F}_{meas}q}
\end{equation}
where $q$ is the detector's quantum efficiency and $\widetilde{F}_{meas}$ is the Fisher information in the measured light field.

To evaluate the Fisher information carried by the outgoing light, we model the reflected and transmitted fields as coherent states with amplitudes $\alpha_R$ and $\alpha_T$, respectively, such that $|\alpha|^2$ gives the photon number. The optical power is then related to the coherent-state amplitude by $P=\hbar \omega_l |\alpha|^2/\tau$, where $\omega_l$ is the probe frequency and $\tau$ is the probe duration. The Fisher information associated with a measurement of the molecular spin projection $J_z$ is therefore

\begin{align}
    \widetilde{F}_{meas}(J_z) &= 4\bigg(\abs{\frac{\partial \alpha_{R}}{\partial S_z}}^2 + \abs{\frac{\partial\alpha_T}{\partial S_z}}^2 \bigg) \nonumber \\
    &= \frac{2\tau}{\hbar\omega}\bigg(\abs{\frac{\partial \mathcal{E}_{R}}{\partial S_z}}^2 + \abs{\frac{\partial \mathcal{E}_T}{\partial S_z}}^2 \bigg) \nonumber \\
    &= \frac{2\tau}{\hbar\omega}(T_o + T_iR_o )\abs{\frac{\partial \mathcal{E}_c}{\partial S_z}}^2 \nonumber \\
    &= \frac{2\tau}{\hbar\omega}(T_o + T_iR_o ) 4\abs{\frac{\partial \mathcal{E}_c}{\partial N_\downarrow}}^2 \nonumber \\
\end{align}

assuming that both the reflected and transmitted fields are measured.

Substituting eq.~\eqref{eq:Ec}, we obtain
\begin{equation}
    \widetilde{F}_{meas} = 16\frac{\abs{\mathcal{E}_{in}}^2}{2}\frac{\tau}{\hbar\omega}\frac{4T_i(T_o + T_iR_o )}{(T_i + T_0)^2}\eta^2\mathcal{L}_a(x_a)\mathcal{T}_0^2
\end{equation}

Combining this result with equations~\eqref{eq:T}, \eqref{eq:S}, and \eqref{eq:ns}, the Fisher information can be expressed in terms of the number of scattered photons as
\begin{equation}
\label{eq:Fmeas}
    \widetilde{F}_{meas} = \frac{16}{N_\downarrow}n_s\eta\mathcal{T}_0  \bigg( 1 - \frac{T_iT_o}{T_i + T_o} \bigg)
\end{equation}

For the case considered here, where $T_o \rightarrow 0$ and $R_o \rightarrow 1$, this expression simplifies to
\begin{equation}
\label{eq:Fmeas_ref}
    \widetilde{F}_{meas} = \frac{16}{N_\downarrow}n_s\eta\mathcal{T}_0
\end{equation}

such that the Fisher information per scattered photon becomes
\begin{equation}
\label{eq:fmeas_per_scattered_photon}
    \widetilde{f}_{meas} = \frac{16}{N_\downarrow}\eta\mathcal{T}_0
\end{equation}

We define the auxiliary quantity $F_{\mathrm{meas}}=\widetilde{F}_{\mathrm{meas}}N/4=\widetilde{F}_{\mathrm{meas}}N_{\downarrow}/2$, which quantifies the metrological gain relative to the standard quantum limit (SQL).

\begin{equation}
    F_{meas} = \bigg( \frac{(\Delta J_z)_{CR}}{(\Delta J_z)_{SQL}} \bigg)^{-2} = 8n_s\eta\mathcal{T}_0
\end{equation}

In the presence of Raman processes, free-space scattering increases the measurement uncertainty~\cite{chen2014cavity}, yielding

\begin{equation}
\label{eq:DeltaJ}
    \frac{(\Delta J_z)^2}{(\Delta J_z)^2_{\mathrm{SQL}}} = \frac{1}{N}\bigg(\frac{4}{n_{s}q\widetilde{f}_{meas}} + 4p_{\mathrm{flip}}n_{s} + p_{\mathrm{loss}}n_{s}\bigg)
\end{equation}
where $(\Delta J_{z})^{2}_{SQL}=N/4$.

\section{III. Estimate of Molecular Ensemble Reuse}

The cavity readout considered here is nearly nondestructive because the probe is operated far from resonance and only a less than one photon is scattered per molecule. For the parameters used in the main text, the optimal operating point corresponds to approximately
\[
n_{\mathrm{sc}}\sim 7\times10^{-3}
\]
free space scattered photons per molecule during a single readout cycle.

Not every scattered photon removes a molecule from the useful spectroscopy manifold. From the branching ratios shown in Fig.~2(a), a molecule initially in $\ket{\downarrow}$ returns to the same state with probability $2/9$, while the remaining population is transferred either to the $N=2$ rotational manifold or to other hyperfine states in $F=1$. In addition, there is a small probability for vibrational branching loss.

If population transferred to $N=2$ and to other vibrational states is treated as irrecoverable, the effective loss probability per scattered photon is approximately
\[
p_{\mathrm{loss}|\gamma}\approx \frac{3}{9}+p_v,
\]
where $p_v$ is the probability for vibrational branching. Taking $p_v\sim 0.02$~\cite{barry2014magneto} gives
\[
p_{\mathrm{loss}|\gamma}\approx 0.35.
\]

The total loss probability per readout cycle is then
\[
p_{\mathrm{loss}}\approx n_{\mathrm{sc}}\,p_{\mathrm{loss}|\gamma}
\approx 0.007\times 0.35
\approx 2.5\times10^{-3}.
\]

The remaining fraction of molecules after $k$ interrogation cycles is therefore
\[
P(k)\approx e^{-k p_{\mathrm{loss}}}.
\]

Defining the useful number of reuses by the condition $P(k)=1/e$ gives
\[
k_{1/e}\approx \frac{1}{p_{\mathrm{loss}}}\sim 400.
\]

In practice, the total number of useful spectroscopy cycles will also depend on the efficiency of state reinitialization and on the trap lifetime. A conservative lower bound is provided by the measured one-body lifetime of optically trapped SrF molecules, $\tau\approx1.3~\mathrm{s}$~\cite{DeMilleSrFCollisions}.  
Because the measured SrF lifetime was obtained in a relatively dense sample where molecule-molecule collisions contributed to trap loss, a dilute cavity-trapped ensemble is expected to exhibit at least comparable, and potentially longer, trapping times.

\section{IV. Results for RaF}
\begin{figure}[H]
    \centering
    \includegraphics[width=0.5\columnwidth]{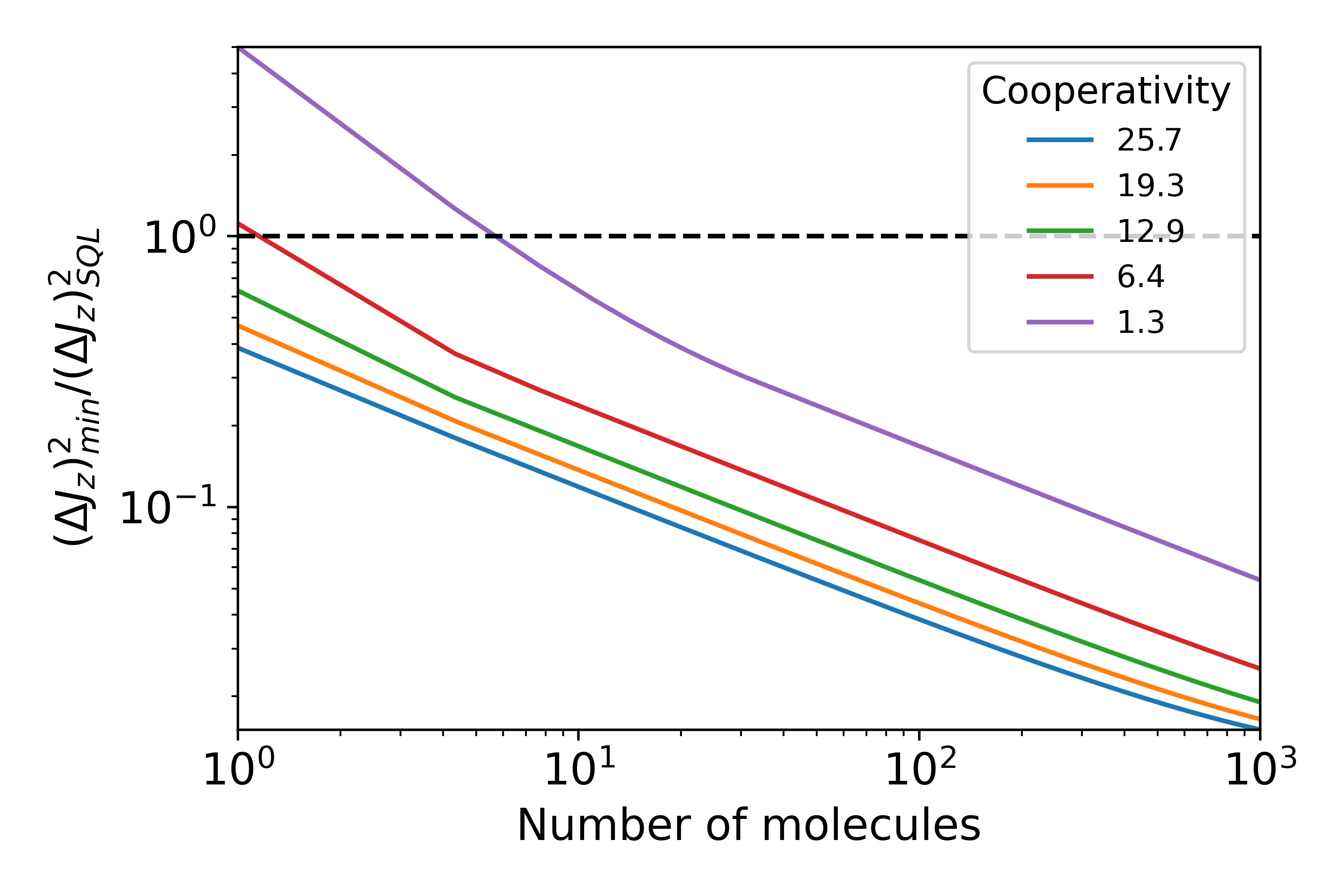}
    \caption{Minimal measurement variance as a function of the number of RaF molecules in the cavity.}
    \label{fig:DN_Nmol_RaF}
\end{figure}

Similarly to SrF, we also estimated the readout uncertainty for RaF as a function of the number of molecules, which can be seen in Fig. \ref{fig:DN_Nmol_RaF}. By comparing to Fig. 2(c) in the main text, we see that the results for RaF and SrF are very similar. This is because, for a far-detuned cavity, the measurement uncertainty mainly depends on the number of molecules, the cooperativity, the quantum efficiency and the loss/flip probabilities defined in Eq. \ref{eq:DeltaJ} \cite{chen2014cavity}. The slight difference between the two figures is due to the fact that the two molecular species have cycling transitions of different wavelengths, $\sim 663$ nm for SrF and $\sim 753$ nm for RaF.

\section{V. Implementation of a Conveyor Belt Magneto-Optical Trap of SrF}

In this section, we present our experimental implementation of a conveyor belt magneto-optical trap (CB MOT) of SrF molecules. The CB MOT is a novel type of molecular MOT which has recently been theoretically proposed~\cite{graceli2025CBMOTTheory} and experimentally realized in several other laser-cooled molecular species~\cite{hallas2026CaOH_CBMOT_published, scarlett2024CaF_CBMOT, zeng2026bafbluemot_published, SrOHODTandCBMOTPublished}. Whereas the first stage of molecular magneto-optical trapping uses light red-detuned from resonance for Doppler cooling, subsequent molecular MOT stages like the CB MOT use light blue-detuned from resonance for sub-Doppler cooling. In a CB MOT, there are two closely spaced laser frequencies which have orthogonal circular polarizations and are blue-detuned from resonance. This scheme has enabled several molecular MOTs to be compressed to sizes $< 100\,\mu\text{m}$~\cite{hallas2026CaOH_CBMOT_published, scarlett2024CaF_CBMOT, SrOHODTandCBMOTPublished} with temperatures typically ranging from $100 - 200\,\mu\text{K}$, making them well-positioned for subsequent high-efficiency loading into optical traps for further applications. 

\begin{figure}
    \centering
    \includegraphics[width=0.45\linewidth]{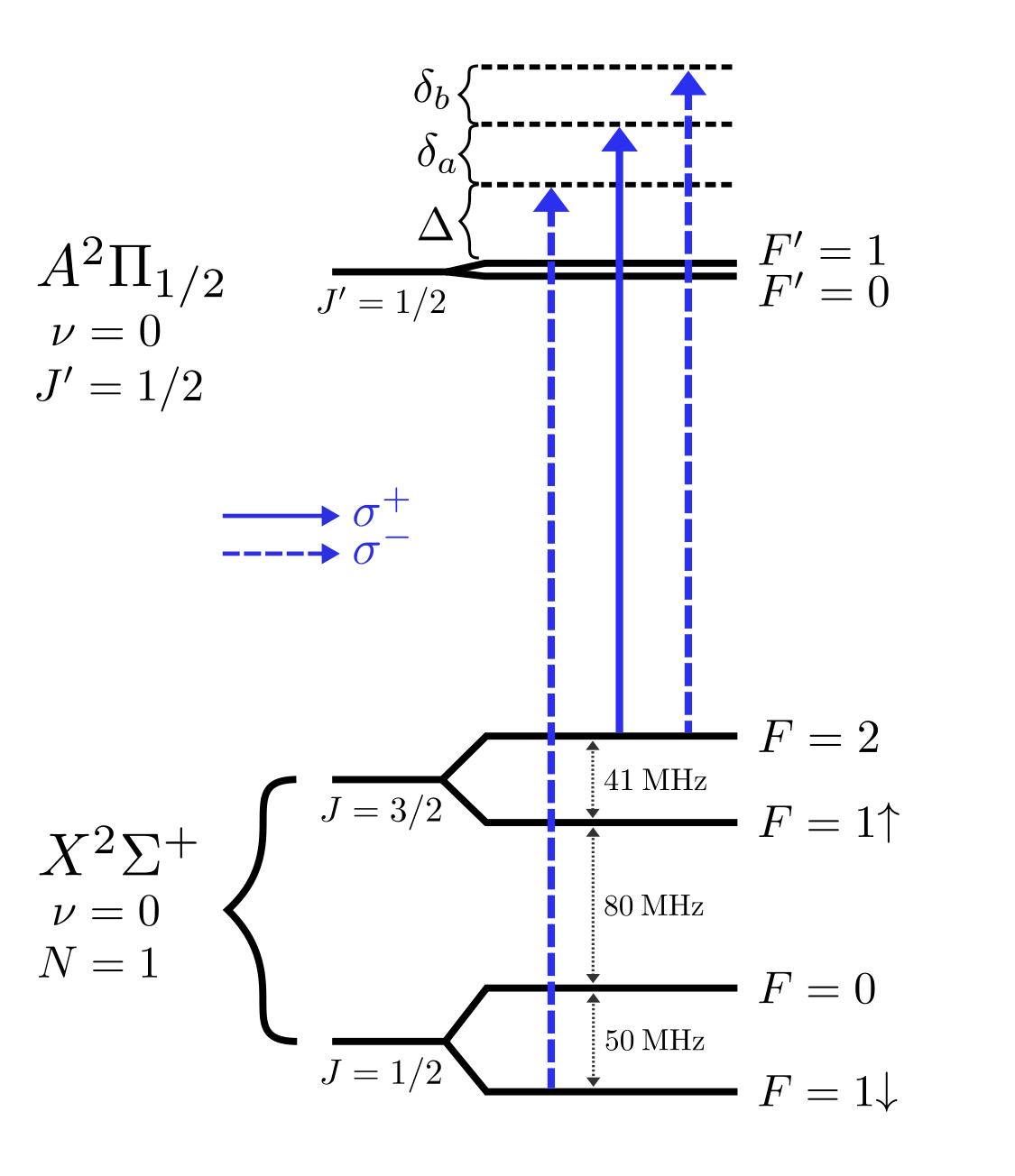}
    \caption{Energy level diagram of relevant states in SrF for the CB MOT scheme. $v$ denotes vibrational quantum number and $N$ denotes rotational quantum number within the electronic states ($X^2\Sigma^+$ and $A^2\Pi_{1/2}$). Hyperfine levels are labeled by $J$ and $F$ quantum numbers, where $\mathbf{J} = \mathbf{N}+\mathbf{S}$ and $\mathbf{F} = \mathbf{I}+\mathbf{J}$, where $\mathbf{S}$ and $\mathbf{I}$ are the electron and nuclear spin angular momentum, respectively. Example circular polarizations ($\sigma^{\pm}$) are given for a laser beam with wave vector $\hat{k}$ pointing towards $+\hat{z}$, and with $\partial B/\partial z > 0$ in that direction. The laser detunings $\Delta$ and $\delta_{a,b}$ are defined in the text. We also apply a weak, near-resonant laser frequency (not shown) to repump from $J=1/2, F=0$.}
    \label{fig:CB MOT level diagram}
\end{figure}

A level diagram showing the frequency scheme for the SrF CB MOT is shown in Fig.~\ref{fig:CB MOT level diagram}. We denote $\Delta$ as the single-photon detuning of all lasers relevant to the CB scheme, $\delta_a$ as the two-photon detuning of the $\sigma^+$ component in the CB, and $\delta_b$ as the two-photon detuning of the $\sigma^-$ component in the CB. Typically, CB MOTs are implemented in a so-called ``1+2" scheme, where one laser is blue-detuned of the lowest hyperfine level ($\ket{F=1\raisebox{0.15ex}{$\downarrow$}}$, see Fig.~\ref{fig:CB MOT level diagram}) while the other two lasers (with orthogonal polarizations with respect to each other) are blue-detuned of the highest hyperfine level ($\ket{F=2}$), providing the actual CB mechanism. 

Our experimental apparatus and procedure for initial laser cooling and magneto-optical trapping of SrF is similar to what was used in previous work~\cite{DeMilleSrFCollisions}. Upgrades to the apparatus are described in Ref.~\cite{QianWangSrFThesis}. For molecular beam slowing and initial magneto-optical trapping, the laser  configuration is essentially the same as in Ref.~\cite{DeMilleSrFCollisions}, though the intensities used for capture into the MOT are higher: for the main cycling transition, we use a per-beam intensity $I_\text{red} \approx 160$ mW/cm$^2$ (effective saturation parameter $s_\text{red} \approx 55$) here, distributed in the ratios $r_\text{red} = [0.63, (0.11, 0.13), 0.06, 0.07]$ for the states $\ket{F=2}, \ket{F=1\raisebox{0.15ex}{$\uparrow$}}, \ket{F=0}$, and $\ket{F=1\raisebox{0.15ex}{$\downarrow$}}$ in the $\ket{N=1}$ manifold, respectively. The parentheses grouping indicates respectively the red and blue-detuned dual-frequency component addressing $\ket{F=1\raisebox{0.15ex}{$\uparrow$}}$. The initial axial magnetic ($B$) field gradient used for capture is $\partial B/\partial z = 18$ G/cm. 

After loading the MOT for 35 ms, we linearly ramp down the laser intensity to $5\%$ of its initial value while simultaneously ramping up the $B$-field gradient to $\partial B/\partial z = 29$ G/cm, over a duration of 25 ms. This results in a compressed MOT with cloud size (Gaussian rms radius) $\sigma \approx 1.1$ mm at temperatures of $T_\text{ax} \approx 1.9(0.1)$ mK, $T_\text{rad} \approx 1.3(0.4)$ mK. Following this stage and 5 ms of hold time to allow equilibration in the MOT, we jump the laser frequencies and intensities to the $\Lambda$-cooling configuration~\cite{DeMilleSrFODT, DeMilleSrFCollisions}. We apply free-space $\Lambda$-cooling on the molecules for 4 ms, which cools the molecular cloud to temperatures as low as $\approx 7(1)\,\mu\text{K}$. This is the coldest reported temperature achieved in a cloud of SrF molecules to-date. 

The $\Lambda$-cooled cloud serves as the starting point for experiments to load a CB MOT of SrF molecules.   Here, we focus on minimizing the size of the resulting molecular cloud, to enable the most efficient subsequent transfer of molecules into an optical cavity for applications such as those described in the main text. We study the dependence of the size of the CB MOT on $\Delta, \delta_a, \delta_b, \partial B/\partial z$, and loading time, and also measure its temperature. 

\begin{figure}
    \centering
    \includegraphics[width=0.85\linewidth]{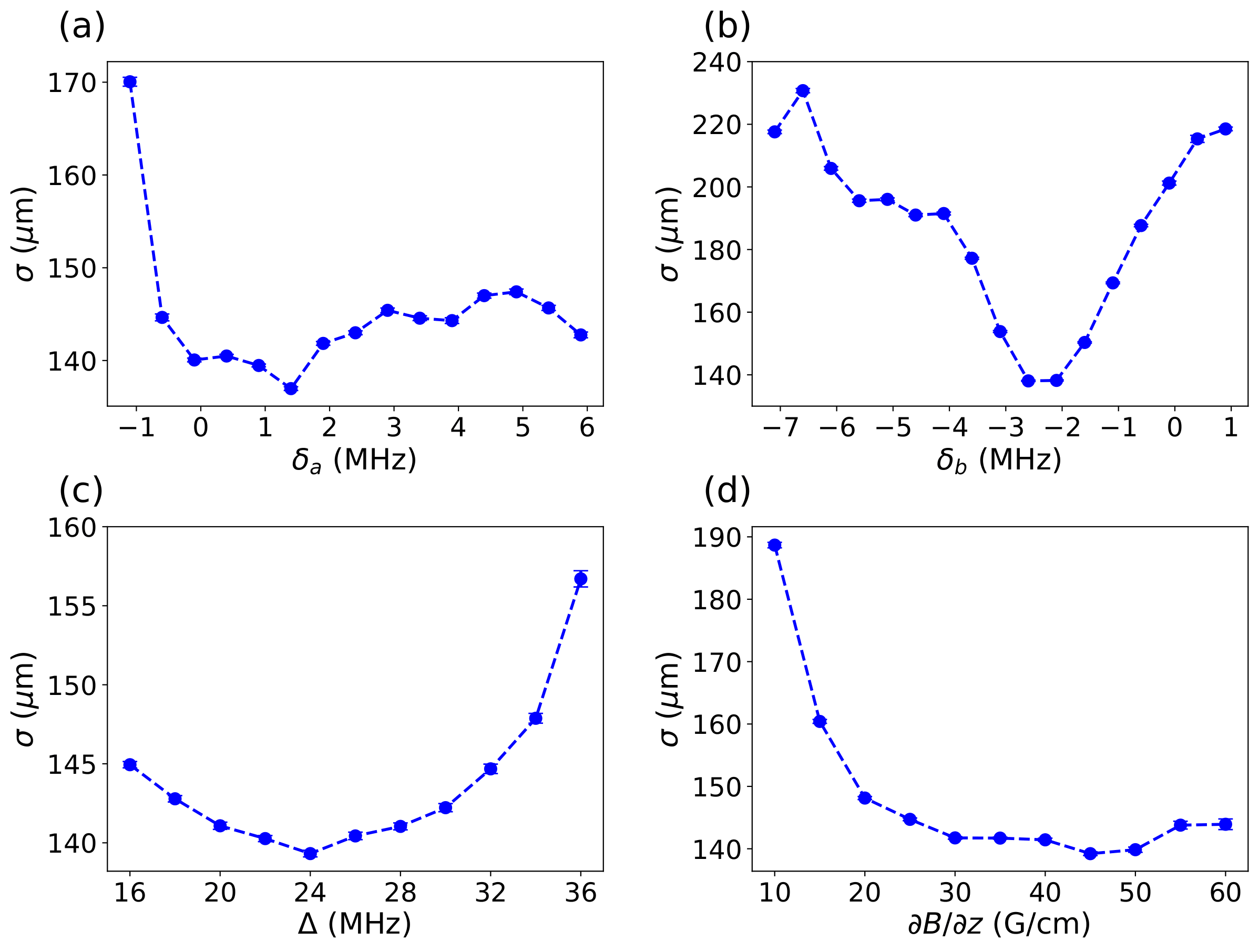}
    \caption{Scans of the SrF CB MOT cloud size as a function of laser frequency detuning and magnetic field gradient. (a-c): MOT cloud size $\sigma$ as a function of $\delta_a$, $\delta_b$, and $\Delta$. (d): MOT cloud size $\sigma$ as a function of B-field gradient. In each individual scan, all non-scanned parameters are held fixed at their optimal values (e.g. $\delta_a = 2\pi \times 1.5$ MHz, $\delta_b = -2\pi \times 2.1$ MHz, $\Delta = 2\pi \times 24$ MHz, and $\partial B/\partial z = 25$ G/cm.) Note that most error bars are smaller than the size of the data points. The curves are merely guides to the eye.}
    \label{fig:CBMOTScans}
\end{figure}

First, we scan the one-photon detuning $\Delta$ to minimize the cloud size (Fig.~\ref{fig:CBMOTScans}(c)). We find that $\sigma$ has a broad minimum around $\Delta = 2\pi \times 24$ MHz. Since the energy splitting between $\ket{F=2}$ and $\ket{F=1\raisebox{0.15ex}{$\uparrow$}}$ is 41.4 MHz, this indicates that the dual frequencies used in the conveyor belt prefer to be tuned closer  to $\ket{F=1\raisebox{0.15ex}{$\uparrow$}}$ than $\ket{F=2}$. This behavior was also observed in the CB MOT of CaF molecules~\cite{scarlett2024CaF_CBMOT}. Next, we individually scan $\delta_a$ and $\delta_b$ (Fig.~\ref{fig:CBMOTScans}(a-b)) while holding $\Delta = 2\pi \times 24$ MHz fixed. We find that optimal trapping in the CB MOT only occurs when $\delta_a > \delta_b$. The optimal values are $\delta_a = 2\pi \times 1.4$ MHz and $\delta_b = -2\pi \times 2.1$ MHz. Both scans indicate sharp increases in $\sigma$ once the frequencies approach the condition $\delta_a \leq \delta_b$. Finally, we scan $\partial B/\partial z$ (Fig.~\ref{fig:CBMOTScans}(d)) while holding $\Delta, \delta_a, \delta_b$ fixed at their optimal values. We find that optimal compression (and capture efficiency) is achieved for $\partial B/\partial z \approx 30$ G/cm. In all of these data, the loading time is 35 ms, and we hold the laser intensities fixed at the following values: for the cycling transition, the per-beam intensity is $I_\text{CB} \approx 60\,\text{mW/cm}^2$ (effective saturation parameter $s_\text{CB} \approx 20$), distributed in the ratios $r_\text{CB} = [0.68, (0.00, 0.23), 0.01, 0.08]$ for the same ordering of states as described above for the red MOT. We observe that the CB MOT cloud size is relatively insensitive to laser intensities over a broad range. Overall, we find that a CB MOT of SrF molecules can reach cloud size as small as $\sigma \approx 140\,\mu\text{m}$. 

\begin{figure}
    \centering
    \includegraphics[width=0.85\linewidth]{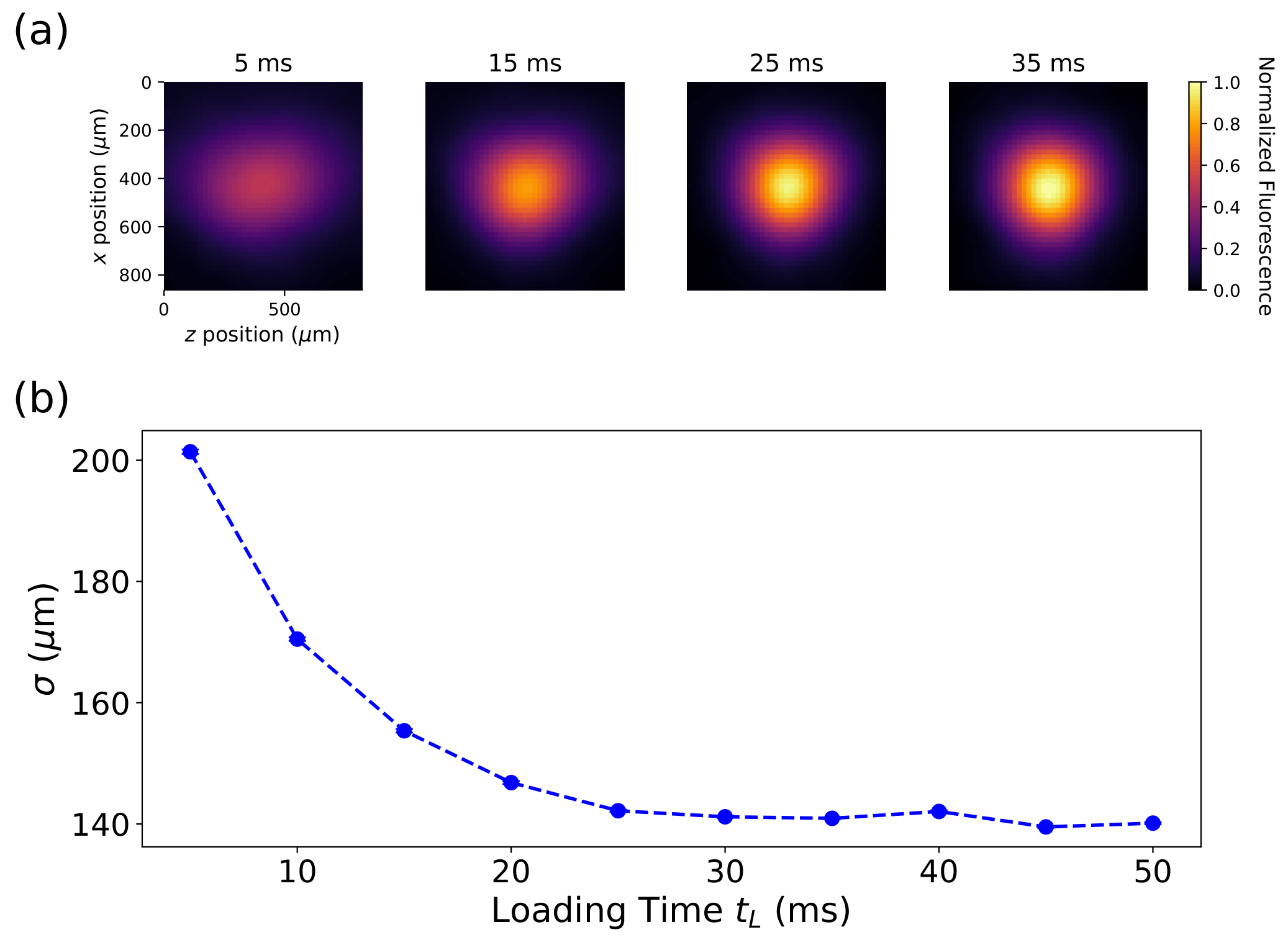}
    \caption{CB MOT size versus loading time. Here, optimal detuning and magnetic field parameters are used: $\delta_a = 2\pi \times 1.5$ MHz, $\delta_b = -2\pi \times 2.1$ MHz, $\Delta = 2\pi \times 24$ MHz, and $\partial B/\partial z = 25$ G/cm. (a) Fluorescence images of the CB MOT at increasing values of $t_L$. For these images, we use 10 ms in-situ exposure to the CB MOT light. (b) Plot of $\sigma$ vs.\ $t_L$. Most error bars are smaller than the size of the data points.}
    \label{fig:LoadingTimeScan}
\end{figure}

We also investigate the behavior of the CB MOT as a function of its loading time, $t_L$. Fig.~\ref{fig:LoadingTimeScan}(b) shows $\sigma$ as a function of $t_L$. The vast majority of compression in size occurs in the first 20 ms, with saturation after roughly 30 ms of loading. We show fluorescence images of the CB MOT in Fig.~\ref{fig:LoadingTimeScan}(a).  
Finally, we measure the temperature of the CB MOT using time-of-flight expansion followed by 5 ms of in-situ fluorescence. We find an axial (radial) temperature $T_\text{ax} = 78(3)\,\mu\text{K}$ ($T_\text{rad} = 63(4)\,\mu\text{K}$). This is the coldest temperature reported to-date in a CB MOT of molecules. 

We point out that the polarizations used for the SrF CB MOT differ from those reported in Ref.~\cite{scarlett2024CaF_CBMOT} and elaborated on in Ref.~\cite{graceli2025CBMOTTheory}; we find that using the polarization scheme reported there leads to no trapping at all. However, our choice is in agreement with the polarizations used in Refs.~\cite{zeng2026bafbluemot_published, LyuAndTarbuttCBMOTExplanation}. This seems to indicate that the proposed mechanism to explain the underlying physics of the CB MOT \cite{graceli2025CBMOTTheory, scarlett2024CaF_CBMOT} may be incorrect. Recently, a different mechanism that explains the behavior we observe was described in Ref.~\cite{LyuAndTarbuttCBMOTExplanation}.

Finally, we compare the SrF CB MOT with our prior result of a ``four-frequency" SrF blue MOT \cite{DeMilleSrFCollisions}. Whereas the SrF CB MOT reaches an average temperature and size of $T_{\rm CB} \approx 70\,\mu$K and $\sigma_{\rm CB}  \approx 140\,\mu$m, respectively, the original SrF blue MOT attained an average temperature and size of $T_b \approx 200\,\mu$K and $\sigma_b \approx 150\,\mu$m, respectively. Hence, for SrF the CB MOT achieves a phase space density $\approx 6$  times larger than the blue MOT, at slightly larger spatial density. The lack of a significant increase in spatial density is surprising, given the larger advantage that the CB MOT affords over alternative blue MOT schemes in all other laser-cooled molecular species~\cite{hallas2026CaOH_CBMOT_published, scarlett2024CaF_CBMOT, zeng2026bafbluemot_published}. Nonetheless, we conclude that the SrF CB MOT should enable efficient loading of molecules into an optical cavity for the future experiments described in the main text.

\section{VI. Quantum State Transfer and Preparation in SrF}

In this section, we discuss experimental protocols and measurements for pure quantum state preparation and transfer between different rotational levels in SrF molecules. This provides a tangible benchmark for efficiencies that one can attain in future experiments  
like those described in the main text. The states proposed for cavity readout are $\ket{\downarrow} = \ket{X^2\Sigma(\nu = 0, N=0, J=1/2, F=0)}$ and $\ket{\uparrow} = \ket{X^2\Sigma(\nu = 0, N=0, J=1/2, F=1, m_F=1)}$. However, the ultracold SrF molecules which will be used to initially load the cavity populate all sublevels in the $N=1$ excited rotational manifold, as a necessity of rotational closure for optical cycling in laser-cooled molecules. 

To prepare molecules in the desired states, we use optical pumping and microwave-driven rotational state transfer on laser-cooled SrF molecules. Similar work has been demonstrated in other laser-cooled molecules such as CaF (e.g. Ref.~\cite{TarbuttCaFMagTrapMWTransfer, DoyleCaFRotationalEntanglement, CheukCaFRotationalEntanglement}) and CaOH (e.g. Ref.~\cite{DoyleCaOHQuantumStateControl}), but ours is the first realization with SrF. In all of the following measurements in this section, our experimental sequence is as follows: starting from a compressed red MOT, we apply free-space lambda cooling \cite{DeMilleSrFODT} and then load a blue MOT \cite{DeMilleSrFCollisions} for sub-Doppler cooling and further compression of the molecular cloud. Immediately after releasing molecules from the MOT, we apply optical pumping and microwaves in succession to the free-space molecular cloud. We then detect the molecular signal with in-situ fluorescence imaging, using red MOT recapture light. 

At the end of any molecular laser cooling sequence, the SrF molecules are distributed across all 12 Zeeman sublevels of the 4 hyperfine states in the $N=1$ rotational manifold. To prepare a pure quantum state, we first optically pump all molecules in $N=1$ into a single Zeeman sublevel, namely the magnetically insensitive $\ket{N=1, J=1/2, F=0}$ hyperfine level. This is accomplished by flashing on the MOT laser frequencies at very low power for 500 $\mu$s, with the exception of the frequency addressing $\ket{N=1, J=1/2, F=0}$, which is turned off. Given the branching ratios for the decay $A^2\Pi_{1/2}, J=1/2 \rightarrow X^2\Sigma, N=1$, it only takes a few scattering events to pump all molecules into the desired state. We achieve a peak optical pumping efficiency of $\approx 94\%$ using this process. Below, we describe how this efficiency was measured.

In the case of SrF, the transition frequency between the $N=1$ and $N=0$ manifolds is roughly 15 GHz~\cite{EricNorrgardPhDThesis, SrFSpectroscopy1970s}. To address this transition, we use an analog signal generator as a coherent microwave source. To achieve arbitrary amplitude and polarization control over the microwaves, we split the output of the signal generator into two distinct transmission lines, place voltage-variable attenuators and voltage-controlled phase shifters along each transmission line, and recombine their output on an orthomode transducer. 
We transmit the microwaves to the molecules using a free-space, high-gain microwave horn coupled to an ellipsoidal focusing mirror. Shim coils aligned to three orthogonal axes in the experiment are used to provide any desired direction of the magnetic field, which defines the quantization axis for the molecules. With our polarization and amplitude control in the microwave system, combined with the fact that microwaves reflect from the metal walls of the vacuum chamber to form a complex pattern of (partial) standing waves, we are able to drive all allowed transitions ($\sigma^{\pm}, \pi$) between the initial $N=1$ state and any sublevel in the $N=0$ manifold.

\begin{figure}
    \centering
    \includegraphics[width=0.85\linewidth]{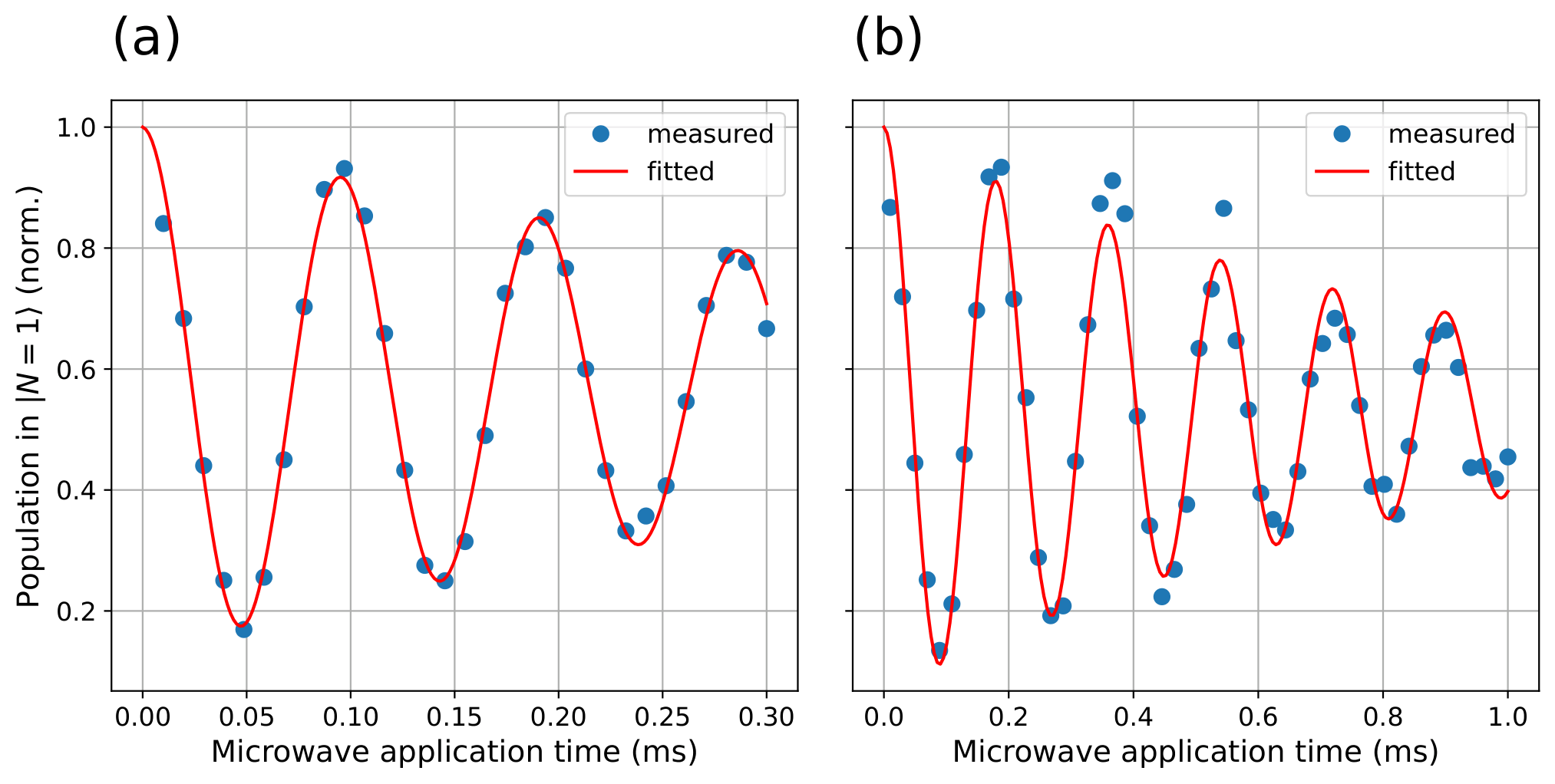}
    \caption{(a) Rabi oscillations between $\ket{N=1, F=0}$ and $\ket{N=0, F=1, m_F=0}$ at a microwave transition frequency of $\omega = 2\pi \times 14890.25$ MHz. (b) Rabi oscillations between $\ket{N=1, F=0}$ and $\ket{N=0, F=0}$ at a microwave transition frequency of $\omega = 2\pi \times 14997.4$ MHz.}
    \label{fig:mw_rabiosc_combined}
\end{figure}

To demonstrate coherent state transfer between $N=1, F=0$ and specific sublevels of the $N=0$ state, we drive resonant transitions. Here, we do not actively control the external field noise or use any dynamical decoupling scheme to extend the coherence time. Hence, the cleanest results come from driving transitions between first-order magnetically insensitive states. We start by driving microwave transitions from $\ket{N=1, J=1/2, F=0}$ to $\ket{N=0, F=1, m_F=0}$.
\footnote{Population can be subsequently transferred to the first-order magnetically sensitive $\ket{\uparrow} = \ket{N=0, F=1, m_F=1}$ state in the cavity QED protocol, via optical pumping.} By applying microwaves for variable amounts of time and measuring the recaptured molecule signal in the MOT, we observe Rabi oscillations between the two states (see Fig.~\ref{fig:mw_rabiosc_combined}(a)).

We extract both the optical pumping efficiency and the microwave transfer efficiency from these data.  We fit the normalized measured molecule signal $p$ in $\ket{N=1}$ as a function of microwave application time $t$ to the function $p = A\cos(\Omega t) \exp(-t/\tau) + C$, where $A$ is the amplitude of the oscillation, $\Omega$ is the Rabi frequency, $\tau$ is the decoherence time, and $C$ is the constant offset in signal. From this fit, we are able to extract the optical pumping efficiency $\epsilon_\text{OP} = 1 - (C-A)/(C+A)$ and the microwave transfer efficiency $\epsilon_\text{MW} = 0.5(1+\exp(-\pi/(\Omega \tau)))$. We find $\epsilon_\text{OP} \approx 87\%$ and $\epsilon_\text{MW} \approx 95\%$.

Next, we repeat the microwave transfer procedure but for the transition from $\ket{N=1, J=1/2, F=0}$ to $\ket{N=0, F=0}$. This is a forbidden transition, so we apply a static magnetic field of up to $B \sim 2$ G using our shim coils to lend transition strength via hyperfine mixing. (The Rabi frequency here is reduced, versus the allowed transition above, by roughly a factor of $\mu_B B/\Delta_\text{hf}$, where $\Delta_\text{hf} \approx 107$ MHz is the energy splitting between hyperfine states in $\ket{N=0}$.) We again observe Rabi oscillations between the two rotational states (see Fig.~\ref{fig:mw_rabiosc_combined}(b)). The extracted efficiencies from these data are $\epsilon_\text{OP} \approx 94\%$ and $\epsilon_\text{MW} \approx 95\%$. 

We remark that in the case of the dipole-allowed transition (Fig.~\ref{fig:mw_rabiosc_combined}(a)), the Rabi flopping decoherence time is roughly $\tau \approx 0.45$ ms, versus $\tau \approx 0.85$ ms in the case of the forbidden transition (Fig.~\ref{fig:mw_rabiosc_combined}(b)). Simulations of the free-space microwave fields inside the vacuum chamber \cite{QianWangSrFThesis}, indicate that the microwaves likely form partial standing waves at the position of the molecular cloud. These would cause microwave intensity gradients across the cloud (since our cloud size is a non-negligible fraction of the microwave wavelength, $\lambda \approx 2$~cm), which in turn would lead to inhomogeneous Rabi frequencies and associated dephasing. In prior testing of microwave-based quantum state transfer \cite{VarunJorapurThesis}, we used a Ramsey sequence on the $\ket{N=1, F=0} \leftrightarrow\ket{N=0, F=1, m_F = 0}$ transition to further study the decoherence mechanisms. We found in that case a Ramsey coherence time of up to $\tau \sim 1.5$ ms, similar to that achieved in rotational transitions in other laser-cooled molecules~\cite{CheukCaFRotationalEntanglement}, in the absence of any mechanism to extend the bare coherence time. 

Our results show clear evidence of controlled population transfer between quantum states of interest in SrF molecules. We have demonstrated single state preparation with efficiency up to $\approx94\%$ and microwave transfer with efficiency up to $\approx95\%$, which is comparable to the highest reported efficiencies in laser-cooled molecules without active internal state error detection and correction~\cite{TarbuttCaFMagTrapMWTransfer,CheukCaFInternalStatePrepEnhanced}. This shows that we can efficiently and coherently transfer SrF molecules with minimal loss from the optical cycling states in $\ket{N=1}$ to the $\ket{\uparrow}$ and $\ket{\downarrow}$ states in $\ket{N=0}$ used in the cavity QED protocols.

\end{document}